\documentclass[preprint,12pt]{elsarticle} 

\usepackage{amssymb}
\usepackage{multirow} 
\usepackage{placeins}
\usepackage{graphicx}
\usepackage{caption}
\usepackage{subcaption}
\usepackage{booktabs}
\usepackage{lscape}

\usepackage{booktabs,caption}
\usepackage[flushleft]{threeparttable}

\begin{document}

\begin{frontmatter}



\title{Effective dose equivalent estimation for humans on Mars}


\author[inst1,inst2]{Miguel Ralha}
\author[inst1,inst3]{Pedro Teles}
\author[inst1,inst2]{Nuno Santos}
\author[inst3]{Daniel Matthiä}
\author[inst3]{Thomas Berger}
\author[inst3]{Marta Cortesão}

\affiliation[inst1]{organization={Departamento de Física e Astronomia, Faculdade de Ciências, Universidade do Porto},
            addressline={Rua do Campo Alegre}, 
            city={Porto},
            postcode={4169-007 }, 
            country={Portugal}}

\affiliation[inst2]{organization={Instituto de Astrofísica e Ciências do Espaço, Universidade do Porto, CAUP},
            addressline={Rua das Estrelas}, 
            city={Porto},
            postcode={4150-762}, 
            country={Portugal}}

\affiliation[inst3]{organization={Centro de Investigação do Instituto Português de Oncologia},
            addressline={Rua Dr. António Bernardino de Almeida}, 
            city={Porto},
            postcode={4200-072}, 
            country={Portugal}}

\affiliation[inst4]{organization={German Aerospace Center, Institute of Aerospace Medicine},
            addressline={Linder Höhe}, 
            city={Cologne},
            postcode={51147}, 
            country={Germany}}

\begin{abstract}
Exposure to cosmic radiation is a major concern in space exploration. On the Martian surface, a complex radiation field is present, formed by a constant influx of galactic cosmic radiation and the secondary particles produced by their interaction with the planet’s atmosphere and regolith. In this work, a Martian environment model was deve\-loped using MCNP6 following the guidelines of the \textit{1st Mars Space Radiation Modeling Workshop}. The accuracy of the model was tested by comparing particle spectra and dose rate results with other model results and measurements from the Radiation Assessment Detector (RAD) onboard the Curiosity rover, taken between November 15, 2015, and January 15, 2016. The ICRP’s voxel-type computational phantoms were then implemented into the code. Organ dose and effective dose equivalent were assessed for the same time period. The viability of a mission on the surface of Mars for extended periods of time under the assumed conditions was here investigated.  

\end{abstract}


\end{frontmatter}




\section{Introduction}
\label{sec:Introduction}

Mars holds a thin atmospheric layer and no appreciable global magnetic field, lacking two of the most important shielding mechanisms that Earth possesses against external sources of radiation. The radiation environment on Mars is mainly caused by the incident galactic cosmic radiation and occasional solar energetic particles, and the subproducts resulting from their interaction with the Martian atmosphere and regolith. Thus, the resulting complex radiation field present at the Red Planet's surface represents a threat to any terrestrial life form due to the increased radiation levels.

Exposures to cosmic radiation are of major concern due to the health detriment they can cause to astronauts. Concerning health risks include cancer, acute radiation sickness, damage to the central nervous system, and degenerative effects, such as cataracts and heart diseases \citep{ED3}. For proposed deep space missions to Mars and beyond, radiation risks from cosmic radiation exposures are likely to be the major limiting factor.

Modern radiation protection programs are based on the basic assumption that risk increases with dose, thus minimising exposures as low as reasonably achievable (ALARA principle). When dealing with low levels of ionising radiation exposure, stochastic health risks to the whole body can be estimated by the effective dose equivalent quantity. An evaluation of organ doses is needed for this quantity, which can then be combined with age and gender-specific coefficients for space-mission’s risk projections. 

The radiation weight factors $w_r$ used for the calculation of the effective dose in terrestrial radiation protection are not recommended to be used for cosmic radiation in space as they are not representative of the average quality factor, specifically for alpha particles and heavier ions. ICRP \citep{ICRP123} recommends using the quality factor $Q$ and the derived quantities of the organ dose equivalent and effective dose equivalent. 

Although the sporadic contributions from Solar Energetic Particles (SEP) can significantly contribute to radiation exposures on the Martian Surface \citep{Intro3,Intro4}, only the exposure from Galactic Cosmic Rays (GCR) are here considered. This work describes the radiation environment present at the Martian surface and the dose quantities required for the estimation of the underlying health risks, using MCNP6 simu\-lations, GCR input spectra from the DLR-model \citep{DLR_3}, and the ICRP voxel-type “Adult Reference Computational Phantoms” \citep{ICRP110}.

\section{MCNP6}
\label{sec:MCNP6}

The simulations were performed with version 6.2 of the Monte Carlo N-Particle (MCNP) transport code \citep{MCNP1}. Following the recommendation of the Los Alamos National Laboratory (LANL) develo\-pers, the inherent default physical models were used for these simulations: the Cascade-Excitation Model (CEM03.03) and, the Los Alamos version of the Quark Gluon String Model (LAQG\-SM03.03) for higher energies (up to TeVs/n).

\section{Determination of the Martian Radiation Environment}
\label{sec:Determination_of_the_Martian_Radiation_Environment}

Since few publications existed in the literature predicting the particle spectra present on the surface of Mars and resulting dose rates, the 1$^{st}$ Mars Space Radiation Modeling Workshop (MSRMW), held in June 2016 in Boulder, CO, USA, brought together as many particle-transport-code groups as possible, working on the calculation of the radiation environment on planetary surfaces, to simulate the radiation environment on the surface of Mars created by the incident GCR \citep{1stMSRMW}. Results from the different models were compared with each other \citep{HASSLER20171, MES_5} and with the measurements performed by the Radiation Assessment Detector (RAD), onboard the Curiosity rover, from the time period between November 15, 2015, and January 15, 2016, to identify the strengths and weaknesses of initial model assumptions and various transport software. 

In this work a model of the Martian environment was developed, GCR propagation through the atmosphere was simulated and particle spectra at the surface level were determined. Additionally, dose rates at the ground level were also assessed. Comparisons of dose rates and particle spectra were performed with results from the MSRMW to gauge the validity of the developed Martian environment model.

\subsection{Model description}
\label{sec:Model_description}

The Martian environment simulation setup was developed by following the workshop’s guidelines \citep{1stMSRMW}:

\begin{enumerate}
    \item The detector mimics the position at the Martian surface corresponding to Curiosity’s landing site on the Gale crater (137.4$^{\circ}$ E longitude, \mbox{4.7$^{\circ}$ S} latitude, -4431 m elevation). 
    \item The Martian atmosphere matches the mean areal density of 23 g/cm$^2$.
    \item The simulation corresponds to solar modulation parameters within the November 15, 2015, to January 15, 2016, time frame. 
    \item Nuclei from the primary galactic cosmic radiation, with atomic numbers from Z = 1 to 28, are used as the external particle source of the simulation.
    \item Particles to be tallied account for all ions up to Nickel (Z = 28), hydrogen and helium isotopes ($^2$H, $^3$H, $^3$He), photons ($\gamma$), neutrons (n), electrons (e$^-$), positrons (e$^+$), positive pions ($\pi^+$), negative pions ($\pi^-$), positive muons ($\mu^+$) and negative muons ($\mu^-$).
    \item Tallies account for particle flux: within a 30$^{\circ}$ off zenith acceptance angle for charged particles only and for all incoming directions (4$\pi$ acceptance) for charged and neutral particles.
    \item Tallies account for average absorbed dose and dose equivalent rates at the Martian surface for all individual incident particles. 
    \item Flux values are reported in specifically prescribed binning structures from 1.04 MeV/n to $9.26 \times 10^5$ MeV/n. 
\end{enumerate}

Additionally, contributors were free to specify the geometry meant to depict the Martian environment, the compositions of the air and regolith, and the models used to reflect the fluctuations in atmospheric density.

The developed geometry was similar to the one described by \citet{RW_MCNP6}: a 6.75 cm radius cylinder running parallel to the $z$-axis extending from 5 m below the regolith’s surface to 100 km above the regolith into the atmosphere. The cylindrical surface was set as a specularly reflective surface to avoid particle loss through the side wall and to simulate the large volume of the Martian atmosphere. The geometry of the atmosphere consisted of 22 layers, having the same material composition but with decreasing density with increasing height. For tallying purposes, a cylindrical detector with 4.5 cm height and 2.25 cm radius, was placed 80 cm above the regolith’s surface—approximately the height of the MSL-RAD inside the Curiosity Rover \citep{MES_1}. 

The Martian atmosphere consists of $\sim$96\% CO$_2$ and trace amounts of other gasses \citep{SAM_1}. The atmospheric altitude density varia\-tion was determined as a function of temperature and pressure with the NASA-Glenn-Research-Center model \citep{NASA_MARS_ATMOS_MODEL}. Martian regolith composition was modelled using the closest Earth-based Martian analogue (JSC Mars-1) \citep{JSC_Mars-1}, but with a mean density of 1.52 g/cm$^3$ set to match the in-situ measurements of the Mars Pathfinder rover. The detector was composed of soft-tissue material, defined according to the element weight fraction composition defined by the ICRP \citep{STDisk}.

The primary particle source placed at the extreme top of the atmospheric column, consisted of a flat, downward oriented disc, allowing all source ions to be generated with a cosine distribution to mimic the isotropic influx of GCR particles. The DLR model \citep{DLR_3} was applied to each source ion (Z = 1 – 28) for the energy spectra determination of each individual ion, for the average time period between November 15, 2015, and January 15, 2016. 

Differential particle spectra at the Martian surface level were calculated from the number of particles crossing a horizontal plane at a given altitude using the F2 tally. For a better assessment of the radiation field, a directio\-nality analysis was performed by integrating the secondary particle spectra over different zenith angle ranges: downward, upward, all directions, and the 30$^\circ$ of zenith angle RAD’s view cone. The absorbed dose and dose equivalent rates were calculated in the soft-tissue disk by recurring to the F6 tally (dose deposition) and to the F6 tally modified with a DF card (automatically multiplies the F6 tally output by the built-in ICRP-60 radiation quality factors). The tallies were allowed over all energy ranges, including those to which the RAD is insensitive, with the purpose of including as much information as possible. In addition to the aforementioned particles to be tallied, one more particle, the neutral pion ($\pi^0$), was included in the simulations due to its important role in high-energy photon production when decaying.

\section{Particle spectra and dose rates results}
\label{sec:Particle_spectra_and_dose_rates_results}

\begin{figure}[h!]
    \makebox[\textwidth][c]{\includegraphics[width=\textwidth]{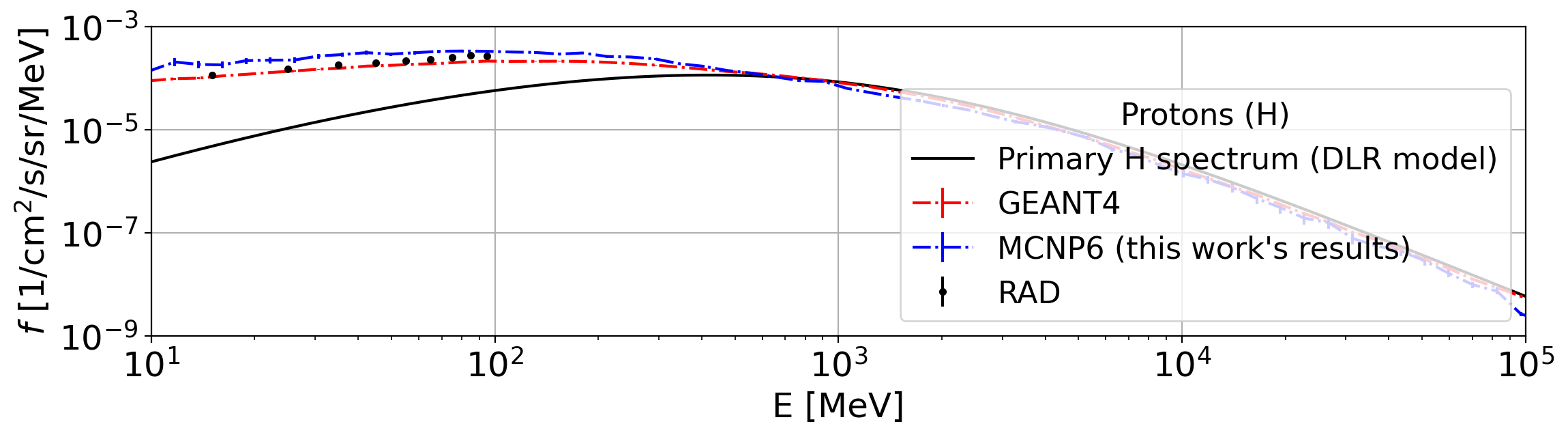}}
    \makebox[\textwidth][c]{\includegraphics[width=\textwidth]{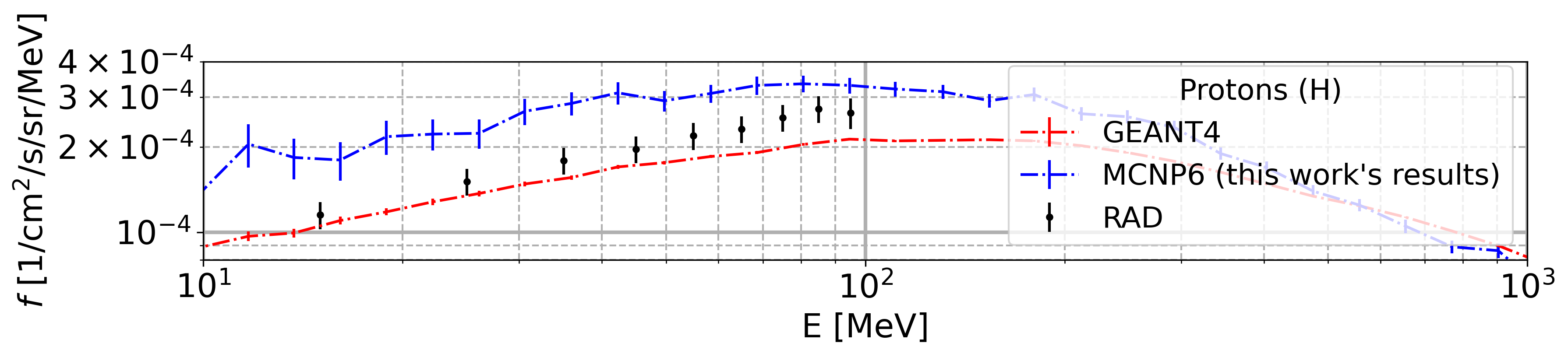}}
    \caption{Proton (H) spectrum measured by RAD on the Martian surface, compared to MCNP6 and GEANT4 results (zenith angle restricted to less than 30$^\circ$). The same data is shown with a different energy range to highlight the RAD data.}
    \label{Spec_H}
\end{figure}

\begin{figure}[h!]
    \makebox[\textwidth][c]{\includegraphics[width=\textwidth]{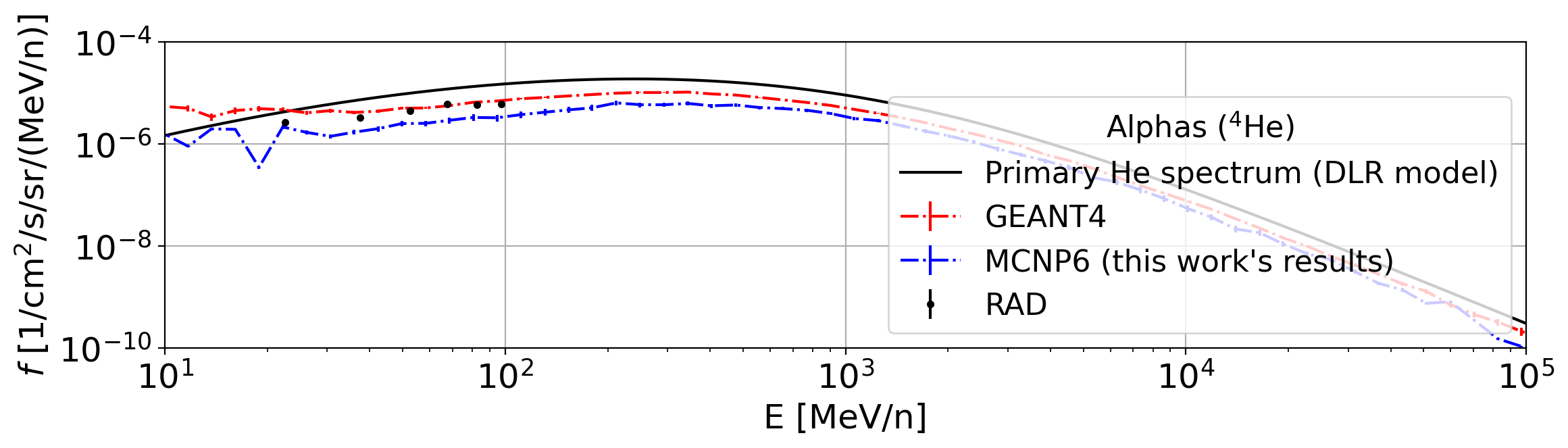}}
    \makebox[\textwidth][c]{\includegraphics[width=\textwidth]{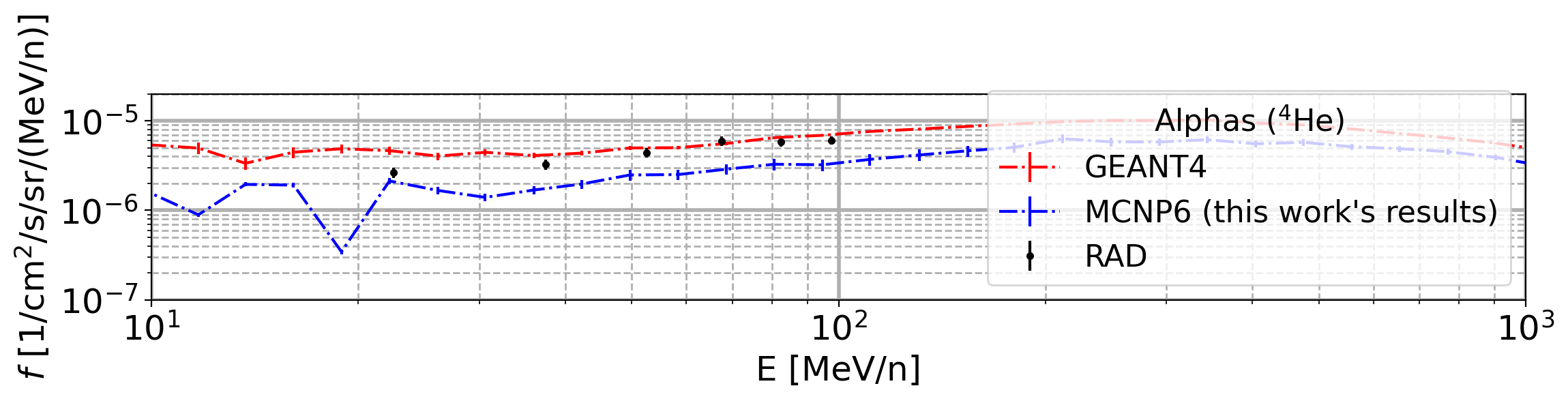}}
    \caption{Helium ion ($^4$He) spectrum measured by RAD on the Martian surface, compared to MCNP6 and GEANT4 results (zenith angle restricted to less than 30$^\circ$). The same data is shown with a different energy range to highlight the RAD data.}
    \label{Spec_He}
\end{figure}

\begin{figure}[h!]
    \makebox[\textwidth][c]{\includegraphics[width=\textwidth]{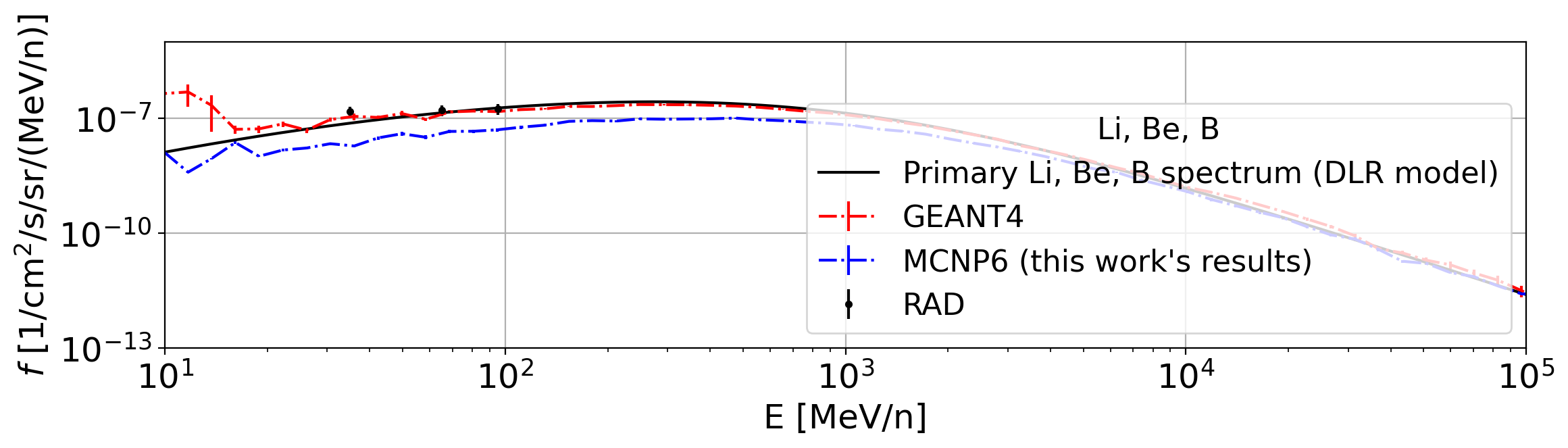}}
    \makebox[\textwidth][c]{\includegraphics[width=\textwidth]{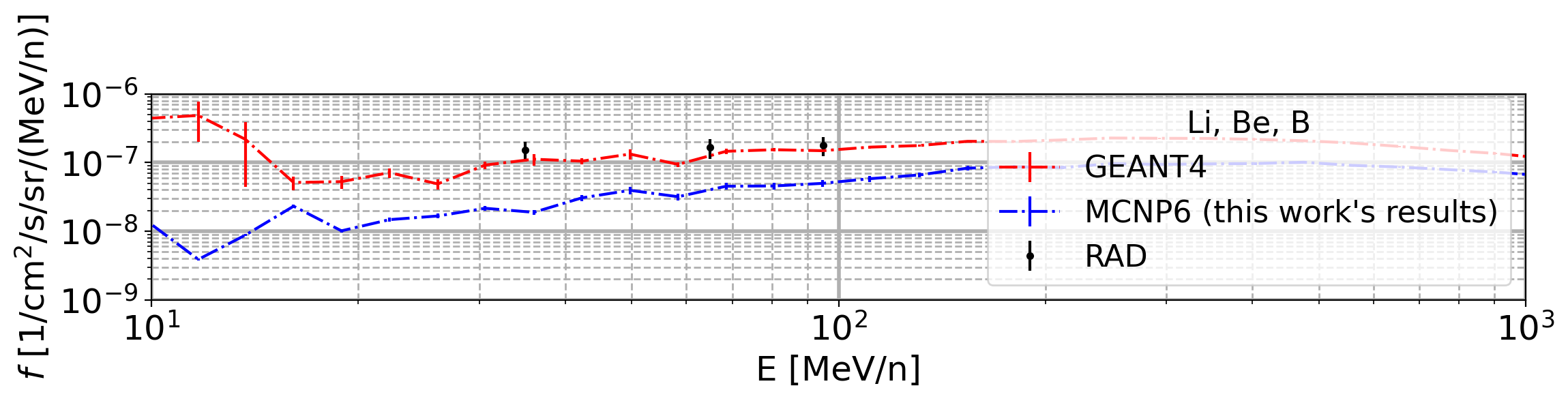}}
    \caption{Lithium, beryllium and boron ion (Z = 3 $-$ 5) spectrum measured by RAD on the Martian surface, compared to MCNP6 and GEANT4 results (zenith angle restricted to less than 30$^\circ$). The same data is shown with a different energy range to highlight the RAD data.}
    \label{Spec_LiBeB}
\end{figure}

\begin{figure}[h!]
    \makebox[\textwidth][c]{\includegraphics[width=\textwidth]{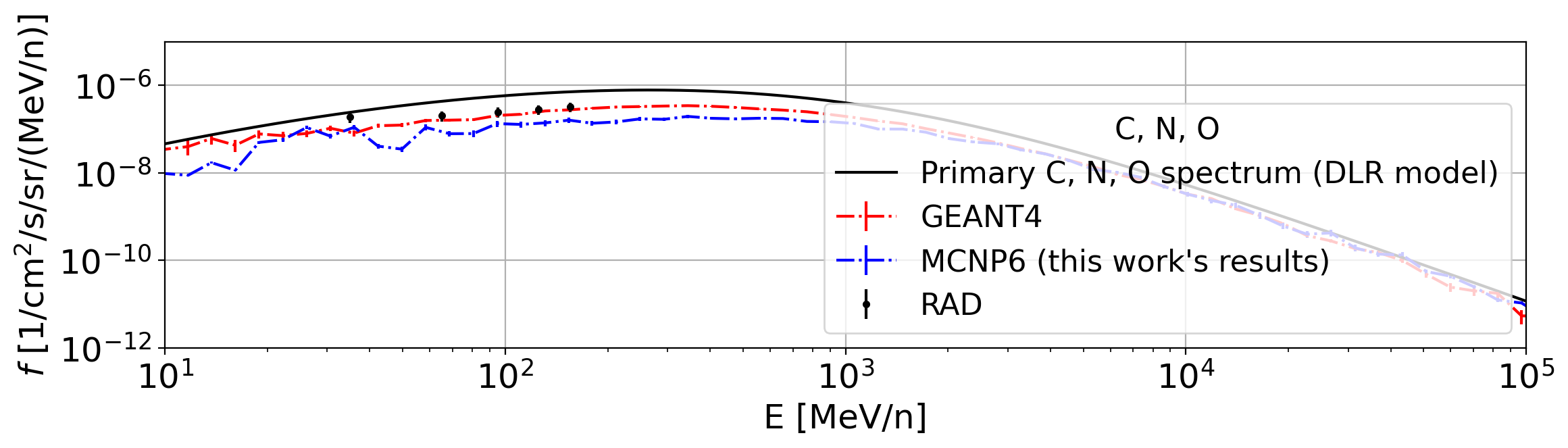}}
    \makebox[\textwidth][c]{\includegraphics[width=\textwidth]{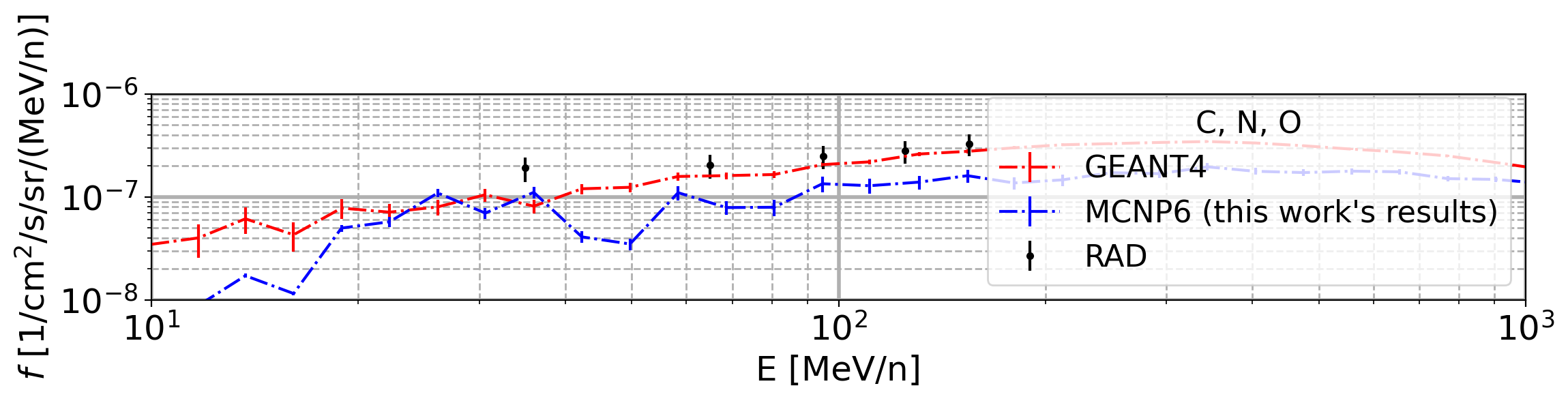}}
    \caption{Carbon, nitrogen and oxygen ion (Z = 6 $-$ 8) spectrum on the Martian surface measured by RAD and compared to model results (zenith angle restricted to less than 30$^\circ$). The same data is shown with a different energy range to highlight the RAD data.}
    \label{Spec_CNO}
\end{figure}

Mean particle fluxes for different solid angle ranges were calculated, namely for the upper hemisphere (downward), lower hemisphere (upward), and for a zenith angle below 30$^\circ$ to mimic RAD’s view cone. MCNP6's calculated spectra were validated by direct comparisons with GEANT4 spectra results, calculated by the German Aerospace Center (DLR) team \citep{MES_7,MES_5}, and RAD’s charged particle measurements \citep{EHRESMANN20173}. Figure \ref{Spec_H} to \ref{Spec_CNO} shows this work’s results for some of the most relevant particles in direct comparison with other relevant data, for zenith angles below 30$^{\circ}$.

\FloatBarrier

\FloatBarrier

At high energies, the slope of the ion spectra closely resembles the one from the primary GCR, pointing to the low atmospheric attenuation that occurs at these high energies. For the proton’s case, in Fig. \ref{Spec_H}, at energies lower than 1 GeV the flux increases above the primary GCR, evidencing the energy loss of a small percentage of primary particles and the production of secondary light particles through nuclear interactions in the atmosphere. With respect to the RAD data, this work’s proton flux is overestimated by $\approx$ 35\% when compared to RAD values, while alpha’s flux values can be found within $\approx$ 60\%. For the heavier nuclei on the surface of Mars (Figs. \ref{Spec_LiBeB} and \ref{Spec_CNO}), the Li, Be, B spectra results present larger deviations distancing around 80\% from RAD measurements. The differences for the C, N, O spectra are lower (around 40\%) presenting a reasonable agreement with other results. As per \citet{MES_5}, the main disparities detected in the spectra of high energy heavy ions can be attributed mostly to variations in source normalisation factors and setup differences among the Monte Carlo codes.

It is important to highlight that for the MSRMW, \citet{MES_7} investigated GEANT4's different physics lists in their simulations, and the values presented here are the ones whose output fitted better to the RAD data. In MCNP6, the use of other physics model combinations, beyond the default models, did not yield better approximations to the RAD measurements.

Additionally, absorbed dose and dose equivalent rates produced by each relevant particle present in the Martian environment were calculated in this work for comparison with RAD’s measurements and other code results, and are shown in Table \ref{table:Dose_rates_ST}. Important to highlight is that neutrons possess unique characteristics in their treatment by the MCNP6 code. In our simulations, we employed the default option for determining the absorbed dose of neutrons, using the F6 tally. By tracking all particles, light and heavy ion recoil particles are produced in the neutron physics.  All interactions are taken into account. However, neutron "heating" (dose deposition) is considered only once per particle production. Additionally, certain reaction products are not generated for neutrons with energies below 20 MeV. According to the MCNP6 manual, this limitation is inherently related to the cross section tables and cannot be trivially fixed \citep{MCNP1}.

From a general comparison between models, this work's absorbed dose and radiation quality factor agreed with other code results, with protons' values being the exception and emerging with relevant discrepancies. From this work’s study, proton’s absorbed dose rates of $178.9 \pm 1.2$ $\mu$Gy/day, were found to be only approximately 10\% lower when compared with the $200.00 \pm 1.04$ $\mu$Gy/day reported by \citet{RW_MCNP6}—who have also used the MCNP6 transport code for the MSRMW—and to be roughly twice as big as other codes’ results, found to be between 85 and 100 $\mu$Gy/day \citep{MES_5}.

In Publication 123 ``Assessment of Radiation Exposure of Astronauts in Space” of the \citet{ICRP123} a set of fluence to organ absorbed dose conversion coefficients and radiation quality factors are provided for the most relevant particles in space: protons, neutrons, charged pions, alpha particles and heavy ions (2 $<$ Z $\leq$ 28). ICRP 123 data sets were combined with the particle spectra determined at the Martian surface to obtain skin dose values. Some relevant particle results found with this new calculation method are also displayed in Table \ref{table:Dose_rates_ST}. Over again, for the absorbed dose rates, significant deviations emerge for protons when compared with this work's MCNP6 results: ICRP 123 method achieved a value of $73.9 \pm 0.3$ $\mu$Gy/day.  

\begin{landscape}
\begin{table}[h!]
\begin{center}
\begin{tabular}{ c c c c c c } 
   \hline
   \multirow{2.5}{*} {\small Particle} & \multicolumn{3}{c}{\small \textbf{MCNP6 (this work's results)}} & \multicolumn{2}{c}{\small \textbf{ICRP 123}} \\
   \cmidrule(rl){2-4} \cmidrule(rl){5-6}
      & \small $D$ ($\mu$Gy/day)  &  \small $Q$ &  \small $H$ ($\mu$Sv/day)  & \small $D$ ($\mu$Gy/day) &  \small $Q$
 \\ 
  \hline
  \footnotesize Protons (H)                  &   \footnotesize 178.9 ($\pm$ \footnotesize 0.7\%)    &  \footnotesize 1.83  ($\pm$ \footnotesize 3.3\%)   &  \footnotesize 327.4 ($\pm$ \footnotesize 0.7\%)   & \footnotesize 73.9 ($\pm$ \footnotesize 0.4\%) &  \footnotesize 1.74 ($\pm$ \footnotesize 0.4\%) \\
  \footnotesize Deuterons ($^2$H)            &   \footnotesize 10.8  ($\pm$ \footnotesize  2.3\%)    &  \footnotesize 1.92  ($\pm$ \footnotesize 3.1\%)   &  \footnotesize 20.8  ($\pm$ \footnotesize 2.3\%)  & — & — \\  
  \footnotesize Tritons ($^3$H)              &   \footnotesize 2.0   ($\pm$ \footnotesize  4.6\%)    &  \footnotesize 2.12  ($\pm$ \footnotesize 6.6\%)   &  \footnotesize 4.2   ($\pm$ \footnotesize  4.5\%) & — & — \\
  \footnotesize Alphas ($^4$He)              &   \footnotesize 14.6  ($\pm$ \footnotesize 2.1\%)    &  \footnotesize 2.03  ($\pm$ \footnotesize 3.0\%)   &  \footnotesize 29.7  ($\pm$ \footnotesize  2.1\%)  & \footnotesize 12.0 ($\pm$ \footnotesize 0.7\%) &  \footnotesize 9.83 ($\pm$ \footnotesize 0.7\%) \\
  \footnotesize Helions ($^3$He)             &   \footnotesize 2.9   ($\pm$ \footnotesize 4.5\%)    &  \footnotesize 2.31  ($\pm$ \footnotesize 6.5\%)   &  \footnotesize 6.7   ($\pm$ \footnotesize 4.5\%)   & — & — \\
  \footnotesize Li, Be, B                    &   \footnotesize 0.9   ($\pm$ \footnotesize 1.2\%)    &  \footnotesize 2.59  ($\pm$ \footnotesize 1.9\%)   &  \footnotesize 2.2   ($\pm$ \footnotesize 1.4\%)   & \footnotesize 0.8 ($\pm$ \footnotesize 1.0\%) & \footnotesize 1.93 ($\pm$ \footnotesize 1.0\%) \\
  \footnotesize C, N, O                      &   \footnotesize 7.1   ($\pm$ \footnotesize 1.0\%)    &  \footnotesize 4.31  ($\pm$ \footnotesize 1.4\%)   &  \footnotesize 30.6  ($\pm$ \footnotesize 1.0\%)   & \footnotesize 6.2 ($\pm$ \footnotesize 1.3\% )&   \footnotesize 2.65 ($\pm$ \footnotesize 1.3\%) \\
 \footnotesize  $Z$ = 9 - 13                 &   \footnotesize 2.5   ($\pm$ \footnotesize 1.6\%)    &  \footnotesize 9.95 ($\pm$ \footnotesize 2.3\%)   &  \footnotesize 24.4  ($\pm$ \footnotesize 1.6\%)    & — & — \\
  \footnotesize $Z$ = 14 - 24                &   \footnotesize 2.3   ($\pm$ \footnotesize 2.6\%)    &  \footnotesize 18.51 ($\pm$ \footnotesize 3.7\%)   &  \footnotesize 42.8  ($\pm$ \footnotesize  2.6\%)  & — & — \\
  \footnotesize $Z$ = 25 - 28                &   \footnotesize 1.5   ($\pm$ \footnotesize 1.4\%)    &  \footnotesize 22.47 ($\pm$ \footnotesize 2.0\%)   &  \footnotesize 32.8  ($\pm$ \footnotesize  1.4\%)  & — & — \\
  \footnotesize Neutrons (n)                 &   \footnotesize 19.7  ($\pm$ \footnotesize 1.2\%)    &  \footnotesize 11.41 ($\pm$ \footnotesize 1.8\%)   &  \footnotesize 225.0 ($\pm$ \footnotesize  1.2\%)  & \footnotesize 20.6 ($\pm$ \footnotesize 0.6\%) &  \footnotesize 4.06 ($\pm$ \footnotesize 0.6\%) \\
  \footnotesize Photons ($\gamma$)           &   \footnotesize 15.1  ($\pm$ \footnotesize 1.2\%)    &  \footnotesize 1.62  ($\pm$ \footnotesize 1.9\%)   &  \footnotesize 24.5  ($\pm$ \footnotesize  1.2\%)  & — & — \\
  \footnotesize Electrons (e$^-$)            &   \footnotesize 63.1  ($\pm$ \footnotesize 2.4\%)    &  \footnotesize 1.0  ($\pm$ \footnotesize 3.0\%)   &  \footnotesize 63.1  ($\pm$ \footnotesize 2.4\%)    & — & — \\
  \footnotesize Positrons (e$^+$)            &  \footnotesize  49.9  ($\pm$ \footnotesize 2.4\%)    &  \footnotesize 1.0  ($\pm$ \footnotesize 3.1\%)   &  \footnotesize 49.9  ($\pm$ \footnotesize 2.4\%)    & — & — \\
  \footnotesize Negative muons ($\mu^-$)     &  \footnotesize  4.5   ($\pm$ \footnotesize 4.0\%)    &  \footnotesize 1.0  ($\pm$ \footnotesize 6.0\%)   &  \footnotesize 4.5   ($\pm$ \footnotesize  4.0\%)   & — & — \\
  \footnotesize Positive muons ($\mu^+$)     &   \footnotesize 5.6   ($\pm$ \footnotesize 3.6\%)    &  \footnotesize 1.0  ($\pm$ \footnotesize 5.0\%)   &  \footnotesize 5.6   ($\pm$ \footnotesize  3.6\%)   & — & — \\
  \footnotesize Negative pions ($\pi^-$)     &   \footnotesize 3.2   ($\pm$ \footnotesize 2.8\%)    &  \footnotesize 1.0  ($\pm$ \footnotesize 3.9\%)   &  \footnotesize 3.2   ($\pm$ \footnotesize  2.8\%)   & — & — \\
  \footnotesize Positive pions ($\pi^+$)     &   \footnotesize 3.5   ($\pm$ \footnotesize 3.1\%)    &  \footnotesize 1.0  ($\pm$ \footnotesize 4.1\%)   &  \footnotesize 3.5   ($\pm$ \footnotesize  3.1\%)   & — & —\\
  \hline
  \footnotesize Total                        &  \footnotesize  388.1 ($\pm$ \footnotesize 0.6\%)    &  \footnotesize 2.32 ($\pm$ \footnotesize 0.8\%)  &  \footnotesize 900.9 ($\pm$ \footnotesize  0.5\%)    & — & — \\
  \hline
\end{tabular}
\end{center}
\caption{Absorbed dose ($D$), radiation quality factor ($Q$) and dose equivalent ($H$) values calculated in this work with MCNP6 (with tally F6) for 4$\pi$ particle incidence, alongside absorbed dose ($D$) and radiation quality factor ($Q$) values, for some relevant particles, calculated with ICRP 123 \citep{ICRP123}. ICRP 123 results were calculated using skin's fluence to organ absorbed dose conversion coefficients and radiation quality factors.} 
\label{table:Dose_rates_ST}
\end{table}
\end{landscape}

Once this work's calculated proton spectra at the Martian surface agrees with other codes' spectra, and the new absorbed dose value—resulting from a combination of particle spectra with ICRP's fluence to absorbed dose conversion coefficients—presents better agreement with other codes reported results, one can only conclude that differences from MCNP6 proton results are linked only with the absorbed dose calculation method (F6 tally) used by the programme. 

An internal communication was initiated with the MCNP6 developers to address and understand these discrepancies (M. \citeauthor{MCNPdev}, private communication, Jan. 4, 2024). The developers have acknowledged the existence of a problem related to tallying protons at high energies. In their response, they conveyed ongoing efforts in their development version MCNP6.3 \citep{MCNP6.3} to rectify issues associated with energy deposition for specific tallies and particles existing in version 6.2. Furthermore, the developers addressed bugs associated with light ion recoil and particle production, speci\-fically linked to the utilization of certain light ion charged particle data files. This acknowledgement and the provided insights have been crucial in understanding and addressing the anomalies observed in proton results obtained in this work.


When compared with the results from \citet{RW_MCNP6}, these work's results show some relevant differences and simi\-larities.  Although it is possible to compare absorbed dose rates calculated with MCNP6 between the two works, caution must be taken when comparing dose equivalent rates data. In their model, \citet{RW_MCNP6} included lower energy cutoffs to mimic RAD’s insensitivity to low-energy particles \citep{MES_1}, which highly influences the outputs of the F6 tallies for the calculated absorbed dose and dose equivalent rates. Exclu\-ding low-energy particles from the simulations means excluding some of the energies corresponding to the higher LET values, thus resulting in a poor determination of the ICRP-60 quality factors. In this work, no such type of low-energy cutoff was enforced achieving seemingly better results for the dose equivalent rates. 

Results suggest that the main contributors to the dose equivalent in the Martian environment are neutrons accounting for $\approx$ 32\%, followed by hydrogen isotopes (protons, deuterons, and tritons) with $\approx$ 23\%, heavier ions with $\approx$ 19\%, electrons and positron with $\approx$ 16\%, helium isotopes ($^3$He, $^4$He) with $\approx$ 5\%, and photons with $\approx$ 3\%. The remaining particles account for less than 3\% of the total dose equivalent.

Table \ref{table:Codes} shows the different model estimates for the absorbed dose rates ($D$), dose equivalent rates ($H$), and average radiation quality factors ($Q$) from the 1$^{st}$ Mars Space Radiation Modeling Workshop as well as the measured values from RAD at the Martian surface \citep{MES_5}.


\begin{table}[h!]
\begin{center}
\begin{tabular}{ c c c c  } 
  \hline
   & \small $D$ ($\mu$Gy/day)  & \small $H$ ($\mu$Sv/day) & \small $Q$
 \\ 
  \hline
  \footnotesize RAD                           &   \footnotesize 233 $\pm$ \footnotesize 12     &   \footnotesize 610 $\pm$ \footnotesize 45   & \footnotesize 2.62 $\pm$ \footnotesize 0.14   \\
  \footnotesize GEANT4 \citep{MES_7}                       &   \footnotesize 206    &  \footnotesize  574   & \footnotesize  2.79  \\
  \footnotesize HZETRN \citep{SLABA2017-2}                       &   \footnotesize 171    &  \footnotesize  536   & \footnotesize  3.14  \\
  \footnotesize MCNP6 \citep{RW_MCNP6}                        &   \footnotesize 307    &  \footnotesize  473   & \footnotesize  1.54  \\
  \footnotesize PHITS \citep{FLORESMCLAUGHLIN2017}                        &   \footnotesize 247    &  \footnotesize  689   &  \footnotesize 2.79  \\
  \hline
\end{tabular}
\end{center}
\caption{Total dose rates ($D$), total dose equivalent rates ($H$) and radiation quality factors ($Q$) calculated with different models and measured by RAD on the surface of Mars for the time period between 15 November 2015, and 15 January 2016 \citep{MES_5}.}
\label{table:Codes}
\end{table}

When proton's absorbed dose rate estimations using fluence to dose conversion coefficients are considered rather than MCNP6 values, the resultant total absorbed dose rate is $283.1 \pm 2.0$ $\mu$Gy/day, the dose equivalent rate $708.7 \pm 3.8$ $\mu$Sv/day and an average quality factor of $2.5 \pm 0.02$ is achieved. Tese results present closeness and good agreement with RAD's measured data, with a relative difference of 21\% for the absorbed dose, 16\% for the dose equivalent, and 4.5\% for the radiation quality factor.

Dose rates are estimated by integrating over a wide range of energy and particle species, depending strongly on the tally types and methods used for absorbed energy estimation. Therefore, comparing dose rates from different models may be subject to larger uncertainties. Also, it strongly depends on the detector’s material and geometry used. For example, this work’s soft tissue detector disk, with a 4.5 cm diameter and height, highly contrasts with the complex geometry of RAD’s detector. As a result, agreement in energy-dependent particle flux is more valuable for model validation than agreement in dose rates from model predictions and measurements, for a given exposure scenario.

\section{Organ Dose and Effective Dose Estimation}
\label{sec:Organ_Dose_and_Effective_Dose_Estimation}

Under current risk acceptance levels for space exploration, astronauts’ space-radiation exposure is of major concern, posing a limiting factor on missions’ duration and viability. Efficient and precise estimations of exposures are essential for improving radiation protection strategies, evaluating health hazards, and minimising harm caused by cosmic radiation during human space exploration. Thus emerges the need to evaluate organ doses through dosimetry and physical considerations, which allows the determination of the stochastic radiation exposure risks. 

\begin{figure}[h!]
   \centering
    \begin{subfigure}[b]{0.46\textwidth}
    \centering
    \includegraphics[width=\textwidth]{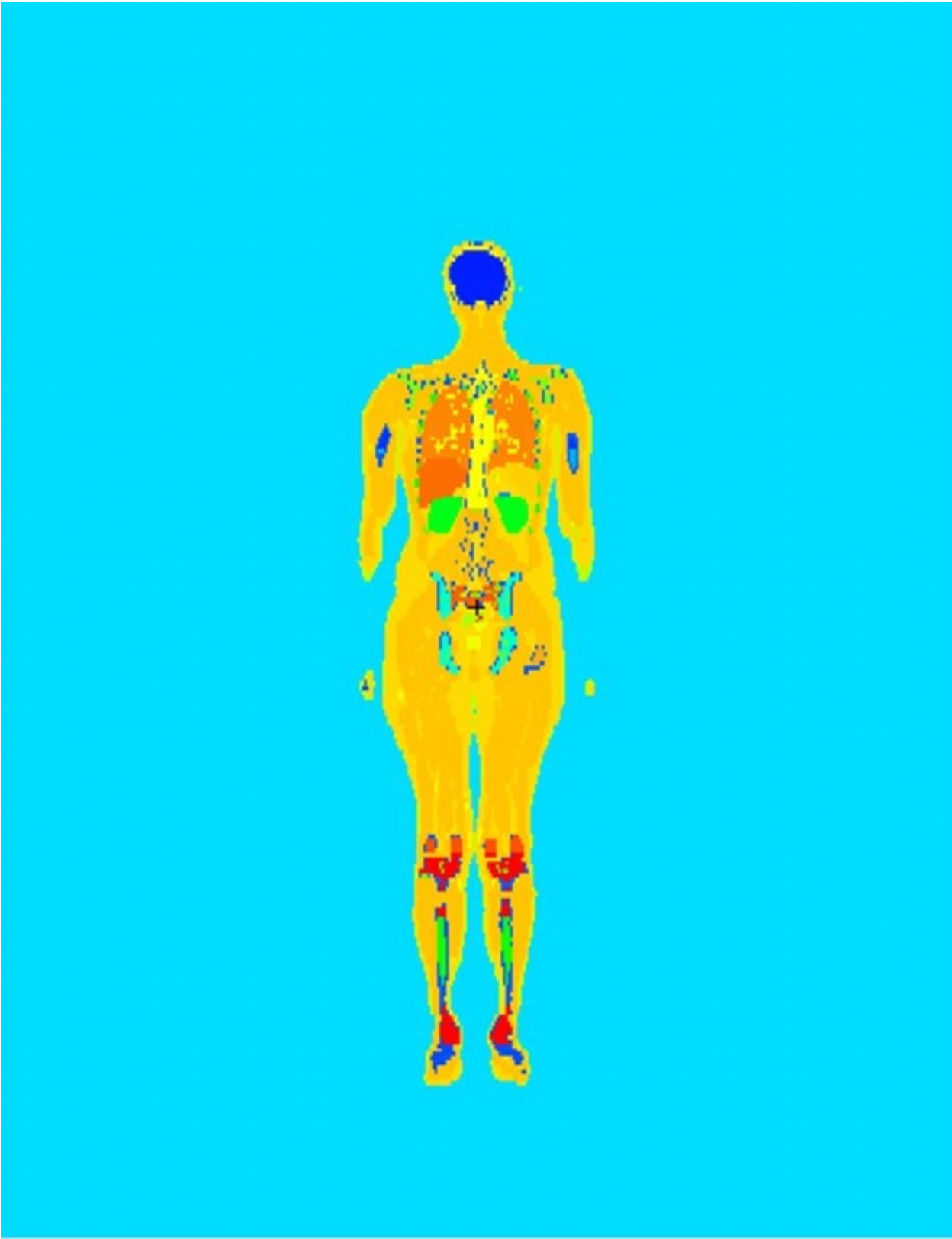}
    \caption{}
    \end{subfigure}
    \begin{subfigure}[b]{0.462\textwidth}
    \centering
    \includegraphics[width=\textwidth]{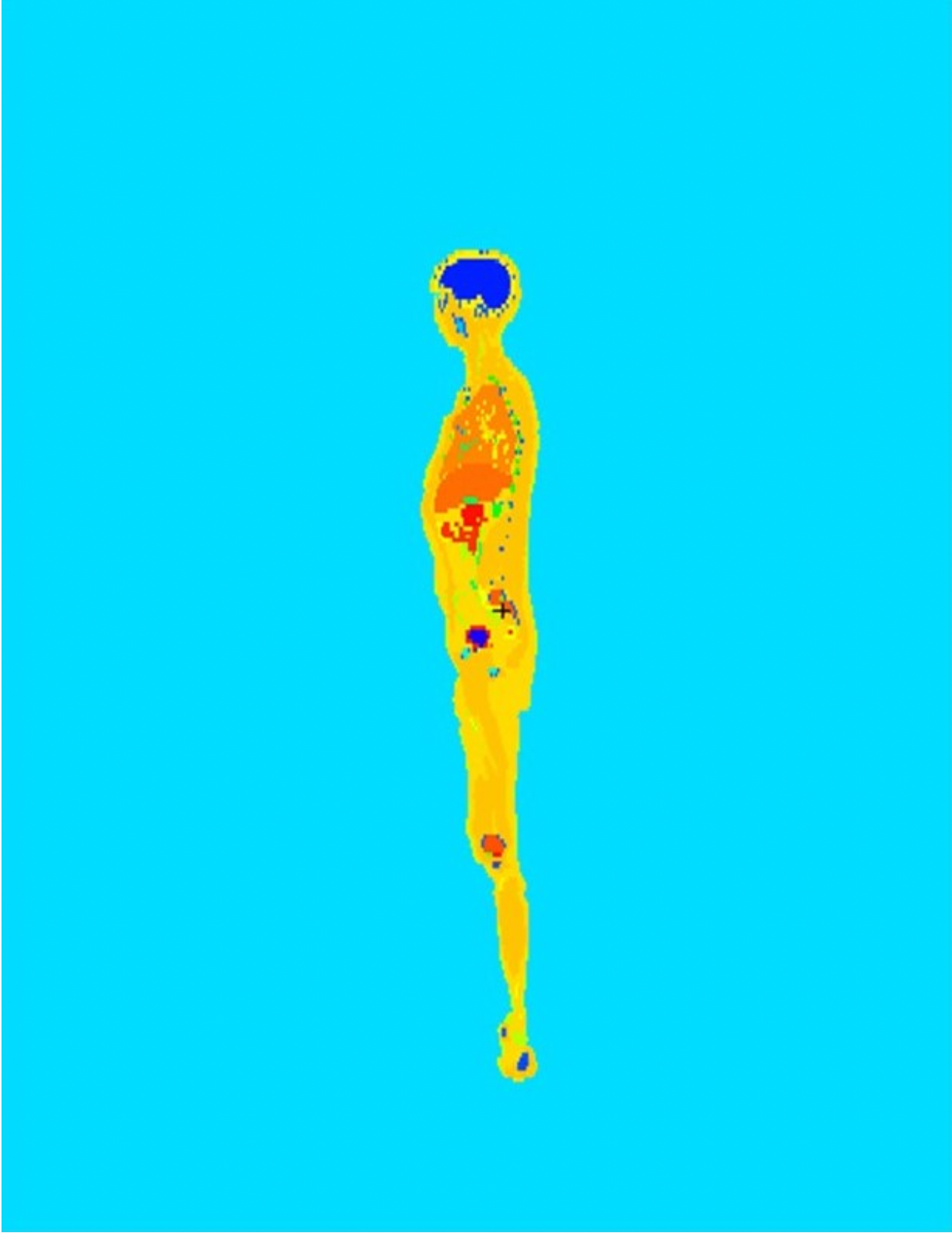} 
    \caption{}
    \end{subfigure}
 \caption{(a) Coronal and (b) sagittal images of the male phantom implemented into the MCNP6 simulations' geometry.}
\label{GeomPhant}
\end{figure}

To that end, in this section, the work performed to simulate an astronaut’s exposure to a radiation field like the one found on the surface of Mars is described. The voxel-type ICRP Adult Reference Phantoms \citep{ICRP110} were implemented into the MCNP6 code (final-result images of the developed geometry in Fig. \ref{GeomPhant}). The main characteristics of the adult male and female ICRP phantoms are summarised in Table \ref{table:1}. Dose rates in the male and female phantoms were calculated in each voxel by irradiating them with the downward and upward-directed particle spectra fields previously calculated in this work.

\begin{table}[h!]
\begin{center}
\begin{threeparttable}
\begin{tabular}{ l l l  } 

  \hline
  Property & Male & Female \\ 
  \hline
  Height (m) & 1.76 & 1.63 \\
  Mass (kg) & 73.0 & 60.0  \\ 
  Number of tissue voxels & 1,946,375 & 3,886,020  \\ 
  Slice thickness (voxel height, mm) & 8.0 & 4.84 \\ 
  Voxel in-plane resolution (mm) & 2.137 & 1.775 \\ 
  Voxel volume (mm$^3$) & 36.54 & 15.25 \\
  Number of columns & 254 & 299 \\ 
  Number of rows & 127 & 137 \\  
  Number of slices & 220 (+2)* &  346 (+2)* \\
  \hline
\end{tabular}
\begin{tablenotes}
\item[*] \footnotesize{Additional skin voxels—not included in the reference height and mass—included at the top of the head and at the bottom of the feet. It is left to the user, and depending on the situation under consideration, whether or not these additional skin voxels are used.}   
\end{tablenotes}
\caption{Main characteristics of the adult male and female reference computational phantoms \citep{ICRP110}.}
\label{table:1}
\end{threeparttable}
\end{center}
\end{table}

Absorbed doses $D_{T,R}$ for each tissue ($T$) and radiation ($R$) type were calculated, as well as the radiation-specific quality factors $Q(\mbox{LET})$. The effective dose equivalent for each sex-type phantom ($H_{E}^{M/F}$) was determined by multiplying the organ/tissue’s dose equivalents $H_T$ by the respective tissue weighting factor $w_T$:

\begin{equation}
    H_{E}^{M/F} = \sum_T w_T \sum_R Q(\mbox{LET}) \cdot D_{T,R} = \sum_T w_T \cdot H_T.
\end{equation}

The ICRP states that mean dose should be given by averaging the organ/tissue’s effective dose equivalent rates from the male ($H^M_E$) and female ($H^F_E$) phantoms: 

\begin{equation}
    H_E = \frac{ 1 }{2} \Bigg(  \sum_T w_T H_{T}^M + \sum_T w_T H_{T}^F \Bigg) = \frac{H_E^M + H_E^F}{2}.
\end{equation}

\begin{table}[b!]
\begin{center}
\begin{tabular}{ c c c c c } 
  \hline
  \small Organ/tissue & \small $D$ ($\mu$Gy/day) & \small $Q$ &  \small $H$ ($\mu$Sv/day) & \small $H_E$ ($\mu$Sv/day)
 \\ 
  \hline
\footnotesize Gonads            &	\footnotesize 121.3	($\pm$	\footnotesize 3.5\%)   &	\footnotesize 2.16 ($\pm$ \footnotesize 5.6\%)	&	\footnotesize 262.0	($\pm$	\footnotesize 4.0\%)	&\\
\footnotesize Urinary bladder  	&	\footnotesize 123.3	($\pm$	\footnotesize 1.0\%)	&	\footnotesize 2.12 ($\pm$ \footnotesize 1.4\%)	&	\footnotesize 261.5	($\pm$	\footnotesize 1.2\%)	    &\\
\footnotesize Oesophagus        &	\footnotesize 169.9	($\pm$	\footnotesize 0.8\%)	&	\footnotesize 2.04 ($\pm$ \footnotesize 1.5\%)	&	\footnotesize 346.6	($\pm$	\footnotesize 1.0\%)	    &\\
\footnotesize Liver	            &	\footnotesize 155.7	($\pm$	\footnotesize 0.2\%)	&	\footnotesize 2.03 ($\pm$ \footnotesize 0.5\%)  &	\footnotesize 316.0	($\pm$	\footnotesize 0.2\%)     &\\
\footnotesize Thyroid           &	\footnotesize 192.5	($\pm$	\footnotesize 1.0\%)	&	\footnotesize 2.07 ($\pm$ \footnotesize 1.4\%)	&	\footnotesize 398.7	($\pm$	\footnotesize 1.2\%)   	&\\
\footnotesize Bone surface	    &	\footnotesize 172.8	($\pm$	\footnotesize 3.5\%)   &	\footnotesize 1.92 ($\pm$ \footnotesize 5.2\%)	&	\footnotesize 331.8	($\pm$	\footnotesize 3.9\%)	&\\
\footnotesize Skin	            &	\footnotesize 201.7	($\pm$	\footnotesize 0.4\%)	&	\footnotesize 2.27 ($\pm$ \footnotesize 0.4\%)	&	\footnotesize 457.4	($\pm$	\footnotesize 0.4\%)	    &\\
\footnotesize Brain	            &	\footnotesize 273.7	($\pm$	\footnotesize 0.2\%)	&	\footnotesize 2.06 ($\pm$ \footnotesize 0.5\%)	&	\footnotesize 564.6	($\pm$	\footnotesize 0.2\%)	              & \footnotesize 343.6 ($\pm$ \footnotesize 0.7\%) \\
\footnotesize Salivary glands	&	\footnotesize 225.7	($\pm$	\footnotesize 1.0\%)	&	\footnotesize 2.08 ($\pm$ \footnotesize 1.4\%)	&	\footnotesize 469.5	($\pm$	\footnotesize 1.0\%)  	&\\
\footnotesize Stomach          	&	\footnotesize 153.7	($\pm$	\footnotesize 0.5\%)	&	\footnotesize 2.04 ($\pm$ \footnotesize 1.0\%)	&	\footnotesize 313.9	($\pm$	\footnotesize 0.6\%)  	&\\
\footnotesize Colon	            &	\footnotesize 140.7	($\pm$	\footnotesize 2.5\%)	&	\footnotesize 2.07 ($\pm$ \footnotesize 3.9\%)	&	\footnotesize 291.0	($\pm$	\footnotesize 2.9\%)	    &\\
\footnotesize Lungs	            &	\footnotesize 183.1	($\pm$	\footnotesize 0.4\%)	&	\footnotesize 1.78 ($\pm$ \footnotesize 0.6\%)	&	\footnotesize 325.0	($\pm$	\footnotesize 0.4\%)   	&\\
\footnotesize Bone marrow(red)  &	\footnotesize 173.2	($\pm$	\footnotesize 3.1\%)	&	\footnotesize 2.17 ($\pm$ \footnotesize 4.6\%)	&	\footnotesize 375.1	($\pm$	\footnotesize 3.4\%)	&\\
\footnotesize Breast	        &	\footnotesize 202.2	($\pm$	\footnotesize 1.7\%)	&	\footnotesize 2.16 ($\pm$ \footnotesize 2.3\%)	&	\footnotesize 437.7	($\pm$	\footnotesize 1.9\%) 	&\\
\footnotesize Remainder tissues	&	\footnotesize 165.3	($\pm$	\footnotesize 0.3\%)   &   \footnotesize 2.14 ($\pm$ \footnotesize 0.5\%)	&	\footnotesize 353.7	($\pm$	\footnotesize 0.4\%) 	&\\
  \hline
\end{tabular}
\end{center}
\caption{Sex-averaged absorbed dose ($D$), dose equivalent ($H$) and effective dose equivalent ($H_E$) values calculated in the ICRP's voxel-type computational phantoms.}
\label{table:Dose_organs}
\end{table}

Absorbed dose, dose equivalent, and effective dose equivalent rates for the Red Planet’s radiation environment are displayed in Table \ref{table:Dose_organs}. Similarly to the previous section, a correction to the protons’ dose rates was employed using ICRP Publication 123 (2013) “Assessment of Radiation Exposure of Astronauts in Space”. The values presented in this table are sex averaged, i.e., represent the average values obtained for the male and female phantoms. The values for the uterus/cervix and the prostate are not sex-averaged, they are sex-specific organs.

 Additionally, Table \ref{table:Dose_RT} presents the detailed values for the remainder tissue constituents.

\begin{table}[h!]
\begin{center}
\begin{tabular}{  c c c c  } 
  \hline
 \small Remainder tissues & \small $D$ ($\mu$Gy/day) & \small $Q$ &  \small $H$ ($\mu$Sv/day) 
 \\ 
  \hline
\footnotesize Adrenals	        &	\footnotesize 147.2	($\pm$	\footnotesize 2.2\%)	&	\footnotesize 2.03 ($\pm$	\footnotesize 3.4\%) &	\footnotesize 299.0	($\pm$	\footnotesize 2.6\%)	\\
\footnotesize ET region	        &	\footnotesize 241.5	($\pm$	\footnotesize 0.9\%)	&	\footnotesize 2.08 ($\pm$	\footnotesize 1.4\%) &	\footnotesize 501.8	($\pm$	\footnotesize 1.1\%)	\\
\footnotesize Gall bladder	    &	\footnotesize 142.9	($\pm$	\footnotesize 1.3\%)	&	\footnotesize 2.04 ($\pm$	\footnotesize 2.0\%) &	\footnotesize 291.3	($\pm$	\footnotesize 1.5\%)	\\
\footnotesize Heart	            &	\footnotesize 163.2	($\pm$	\footnotesize 0.2\%)	&	\footnotesize 2.03 ($\pm$	\footnotesize 0.5\%) &	\footnotesize 331.3	($\pm$	\footnotesize 0.3\%)	\\
\footnotesize Kidneys	        &	\footnotesize 147.0	($\pm$	\footnotesize 2.1\%)	&	\footnotesize 2.03 ($\pm$	\footnotesize 3.4\%) &	\footnotesize 298.6	($\pm$	\footnotesize 2.8\%)	\\
\footnotesize Lymphatic nodes	&	\footnotesize 151.4	($\pm$	\footnotesize 2.7\%)	&	\footnotesize 2.08 ($\pm$	\footnotesize 3.8\%) &	\footnotesize 331.4	($\pm$	\footnotesize 2.9\%)	\\
\footnotesize Muscles	        &	\footnotesize 166.5	($\pm$	\footnotesize 0.2\%)	&	\footnotesize 2.15 ($\pm$	\footnotesize 0.5\%) &	\footnotesize 357.2	($\pm$	\footnotesize 0.2\%)	\\
\footnotesize Oral mucosa	    &	\footnotesize 220.1	($\pm$	\footnotesize 1.0\%)    &	\footnotesize 2.06 ($\pm$	\footnotesize 1.5\%) &	\footnotesize 452.5	($\pm$	\footnotesize 1.3\%)	\\
\footnotesize Pancreas	        &	\footnotesize 137.9	($\pm$	\footnotesize 0.4\%)	&	\footnotesize 2.04 ($\pm$	\footnotesize 0.5\%) & \footnotesize 281.5	($\pm$	\footnotesize 0.5\%)	\\
\footnotesize Small intestine	        &	\footnotesize 134.6	($\pm$	\footnotesize 0.2\%)	&	\footnotesize 2.07 ($\pm$	\footnotesize 0.5\%) &	\footnotesize 278.0	($\pm$	\footnotesize 0.2\%)	\\
\footnotesize Spleen	        &	\footnotesize 165.8	($\pm$	\footnotesize 0.4\%)	&	\footnotesize 2.02 ($\pm$	\footnotesize 0.5\%) &	\footnotesize 334.3	($\pm$	\footnotesize 0.4\%)	\\
\footnotesize Thymus	        &	\footnotesize 178.5	($\pm$	\footnotesize 0.6\%)    &	\footnotesize 2.06 ($\pm$	\footnotesize 1.0\%) &	\footnotesize 368.1	($\pm$	\footnotesize 0.7\%)	\\
\footnotesize Uterus/cervix  	&	\footnotesize 125.0	($\pm$	\footnotesize 0.8\%)	&	\footnotesize 2.10 ($\pm$	\footnotesize 1.4\%) &	\footnotesize 262.9	($\pm$	\footnotesize 1.1\%)	\\
\footnotesize Prostate	        &	\footnotesize 113.5	($\pm$	\footnotesize 1.3\%)    &	\footnotesize 2.12 ($\pm$	\footnotesize 1.9\%) &	\footnotesize 240.6	($\pm$	\footnotesize 1.4\%)  \\
\hline
\footnotesize Total 	        &	\footnotesize 165.3	($\pm$	\footnotesize 0.3\%) &   \footnotesize 2.14 ($\pm$   \footnotesize 0.5\%)	&	\footnotesize 353.7	($\pm$	\footnotesize 0.4\%)\\
\hline
\end{tabular}
\end{center}
\caption{Absorbed dose ($D$) and dose equivalent ($H$) in the remainder tissues of the ICRP's voxel-type computational phantoms.}
\label{table:Dose_RT}
\end{table}

The dominance of downward-directed particles results in more irradiated upper-body organs, namely the brain, the ET region, the salivary glands, the oral mucosa, the breasts, and the thyroid. Additionally, natural shielding is provided by these upper-body and more superficial structures, allowing only more penetrating radiation to reach the lower-body organs. This can be seen in the prostate, gonads, uterus/cervix, urinary bladder, small intestine, etc., which show lower dose rates. 

Regarding the radiation quality factors, two interesting values contrast: the highest value from the skin ($Q = 2.27 \pm 0.01$) and the lowest from the lungs ($Q = 1.78 \pm 0.01$). Because of their higher quality factors at small to medium shielding depths, heavy ions appear to represent a big portion of the overall dose equivalent rate value, as suggested by the skin's $Q$ factor. Consequently, the skin emerges as an important layer of protection from low penetrating radiation. For the lungs, bronchioles in the ICRP’s voxel-type computational phantoms are represented as homogeneous tissue with their density averaged between the higher-density bronchiolar tissue and the included air. This unrealistic low-density structure ($\rho = 0.385$ g/cm$^3$) directly results in a lower calculated stopping power for the charged particles in the medium. Caution must be taken when using the lung’s calculated dose equi\-valent values once the radiation quality factor is probably underestimated. 

Results indicate that, on average, a total of $343.6 \pm 2.4$ $\mu$Sv/day would be the effective dose equivalent received by an astronaut present on the pla\-netary surface—achieving $H_{E}^{M} = 339.3 \pm 2.4$ and $H_{E}^{F} = 348.0 \pm 2.4$ in the male and female phantoms, respectively. An additional study was performed to assess the contributions of each particle type to the total effective dose equivalent, with results being shown in Table \ref{table:Particle_Dose}. Once the radiation field can be divided into two components according to the radiation directionality, the contributions from downward- ($D$) and upward-directed ($U$) particles were studied separately. The total field (4$\pi$ irradiation) is also displayed, resulting from the direct sum of the two components.

\begin{table}[h!]
\begin{center}
\begin{tabular}{ c c c c } 
  \hline
  \small Particle & \small $H_E^D$ ($\mu$Sv/day) & \small $H_E^U$ ($\mu$Sv/day) & \small  $H_E^{4\pi}$ ($\mu$Sv/day)  \\ 
  \hline
  \footnotesize Protons (H)              &	\footnotesize 111.9	($\pm$	\footnotesize 1.9\%)	&	\footnotesize 7.0 	  ($\pm$	\footnotesize 1.3\%)	 &	\footnotesize 118.9	    ($\pm$	\footnotesize 1.8\%)	\\
  \footnotesize Deuterons ($^2$H)        &	\footnotesize 1.9	($\pm$	\footnotesize 1.1\%)	&	\footnotesize 0.3	  ($\pm$	\footnotesize 0.7\%)    &	\footnotesize 2.2 	    ($\pm$	\footnotesize 0.9\%)	\\
  \footnotesize Tritons ($^3$H)          &	\footnotesize 0.4 	($\pm$	\footnotesize 0.7\%)    &	\footnotesize 0.01    ($\pm$	\footnotesize 0.7\%)    &	\footnotesize 0.4 	    ($\pm$	\footnotesize 0.7\%) \\
  \footnotesize Alphas ($^4$He)          &	\footnotesize 16.2	($\pm$	\footnotesize 1.2\%)	&	\footnotesize 0.01    ($\pm$	\footnotesize 0.9\%)    &	\footnotesize 16.2 	    ($\pm$	\footnotesize 1.2\%)	\\
  \footnotesize Helions ($^3$He)         &	\footnotesize 1.0   ($\pm$	\footnotesize 1.0\%)	&	\footnotesize 0.1     ($\pm$	\footnotesize 0.1\%)    &	\footnotesize 1.1 	    ($\pm$	\footnotesize 0.9\%)	\\
  \footnotesize Li, Be, B                &	\footnotesize 0.7   ($\pm$	\footnotesize 1.5\%)	&	\footnotesize 0.0 	  	                                &	\footnotesize 0.7 	    ($\pm$	\footnotesize 1.5\%)	\\
  \footnotesize C, N, O                  &	\footnotesize 8.4   ($\pm$	\footnotesize 2.0\%)	&	\footnotesize 0.0 	                                	 &	\footnotesize 8.4 	    ($\pm$	\footnotesize 2.0\%)	\\
  \footnotesize $Z$ = 9 - 13             &	\footnotesize 6.0 	($\pm$	\footnotesize 2.0\%)	&	\footnotesize 0.0     	                                 &	\footnotesize 6.0 	    ($\pm$	\footnotesize 2.0\%)	\\
  \footnotesize $Z$ = 14 - 24            &	\footnotesize 9.6   ($\pm$	\footnotesize 5.1\%)	&	\footnotesize 0.0                                    	 &	\footnotesize 9.6 	    ($\pm$	\footnotesize 5.1\%)	\\
  \footnotesize $Z$ = 25 - 28            &	\footnotesize 7.0  	($\pm$	\footnotesize 1.6\%)	&	\footnotesize 0.0                                     	 &	\footnotesize 7.0 	    ($\pm$	\footnotesize 1.6\%)	\\
  \footnotesize Neutrons (n)             &	\footnotesize 56.8	($\pm$	\footnotesize 1.0\%)	&	\footnotesize 43.3	  ($\pm$	\footnotesize 1.2\%)	 &	\footnotesize 100.1 	($\pm$	\footnotesize 0.8\%)	\\
  \footnotesize Photons ($\gamma$)       &	\footnotesize 8.5	($\pm$	\footnotesize 1.8\%)	&	\footnotesize 1.3 	  ($\pm$	\footnotesize 2.3\%)	 &	\footnotesize 9.7 	    ($\pm$	\footnotesize 1.5\%)	\\
  \footnotesize Electrons (e$^-$)        &	\footnotesize 24.7  ($\pm$	\footnotesize 1.2\%)	&	\footnotesize 3.5 	  ($\pm$	\footnotesize 1.1\%)	 &	\footnotesize 28.2	    ($\pm$	\footnotesize 1.1\%)	\\
  \footnotesize Positrons (e$^+$)        &	\footnotesize 23.4  ($\pm$	\footnotesize 1.2\%)	&	\footnotesize 2.5 	  ($\pm$	\footnotesize 1.2\%)	 &	\footnotesize 25.9	    ($\pm$	\footnotesize 1.1\%)	\\
  \footnotesize Negative muons ($\mu^-$) &	\footnotesize 2.4  	($\pm$	\footnotesize 1.3\%)	&	\footnotesize 0.6 	  ($\pm$	\footnotesize 1.7\%)	 &	\footnotesize 3.0 	    ($\pm$	\footnotesize 1.0\%)	\\
  \footnotesize Positive muons ($\mu^+$) &	\footnotesize 3.0  	($\pm$	\footnotesize 1.3\%)	&	\footnotesize 0.7 	  ($\pm$	\footnotesize 1.4\%)	 &	\footnotesize 3.8 	    ($\pm$	\footnotesize 1.1\%)	\\
  \footnotesize Negative pions ($\pi^-$) &	\footnotesize 0.1 	($\pm$	\footnotesize 0.9\%)    &	\footnotesize 1.1     ($\pm$	\footnotesize 0.9\%)	 &	\footnotesize 1.2	    ($\pm$	\footnotesize 0.8\%)	\\
  \footnotesize Positive pions ($\pi^+$) &	\footnotesize 0.1 	($\pm$	\footnotesize 0.9\%)    &	\footnotesize 1.1     ($\pm$	\footnotesize 0.9\%)	 &	\footnotesize 1.2 	    ($\pm$	\footnotesize 0.8\%)	\\
  \hline													
  \footnotesize Total                    &	\footnotesize 282.1	($\pm$	\footnotesize 0.8\%)   &	\footnotesize 61.50	  ($\pm$	\footnotesize 0.8\%)	     &	\footnotesize 343.6 	($\pm$	\footnotesize 0.7\%)	\\
 \hline	
\end{tabular}
\end{center}
\caption{Particle-type and radiation directionality (upward and downward) contributions to the total effective dose equivalent ($4\pi$ incidence).}
\label{table:Particle_Dose}
\end{table}

It was found that from the overall field, the downward-directed component comprises 82\% of the total effective dose equivalent, while the upward component only represents 18\%.

Particles present on the primary GCR ($1 <$ Z $< 28$) represent $\approx$ 50\% of the total effective dose equivalent, with the protons’ field being responsible for the majority of the energy imparted to the phantoms ($\approx$ 35\%). As the atomic number increases, the contribution to the upward component by heavy ions rapidly diminishes, being negligible for Z $\geq$ 3. 

Neutron, electron, and positron contributions to the effective dose equi\-valent are also significant, comprising 45\% of the total value, with neutrons alone comprising $\approx$ 29\%. Neutron contributions from the downward and upward components are close, due to the near-isotropy regime found in the field for energies lower than 100 MeV. For electron and positron contributions, a more pronounced difference was expected due to the differences found in the overall fields from both particles. Regarding pions, contributions are only significant for the upward component, due to the strong production of these particle types in the Martian regolith. 

In our initial calculations, we did not account for the shielding effects of spacesuits and habitats, focusing instead on the worst-case scenario of unshielded exposure. However, it is important to consider the potential impact of these shielding elements. For instance, a spacesuit with an aluminum-equivalent shielding thickness ranging from 0.3 g/cm$^2$ (light spacesuit) to 1 g/cm$^2$ (heavy spacesuit), would contribute less than 5\% additional shielding compared to the minimum vertical atmospheric shielding on Mars (over all zenith angles effective shielding is much higher). This minimal increase indicates that the spacesuit's contribution to overall shielding is relatively minor, as further confirmed by the works developed by \citet{SLABA2017} and \citet{BALLARINI2006}. Therefore, the omission of spacesuit shielding in our initial calculations represents a second-order effect and does not significantly alter the overall GCR dose assessments. Future studies will, nonetheless, include detailed shielding models for more precise dose calculations.

\subsection{Dose limits for astronauts}
\label{sec:Dose_limits_for_astronauts}

Space agencies impose career dose limits, intended to limit the increased risk of cancer to an acceptable level. Due to differences in tissue types and sensitivities, latency effects, and differences in average life span between genders, the relationship between radiation exposure and risk is age- and gender-specific \citep{ED4}. Following this, space agencies have developed individual approaches to limit definitions, to account for these variables. 

The National Aeronautics and Space Administration (NASA) developed a risk-based system for radiation protection that limits individual occupational radiation exposures to a lifetime 3\% excess risk of cancer death, within a 95\% confidence interval (CI) for a 1-year mission \citep{ED5}. The NASA Space Cancer Risk (NSCR) model NSCR-2012 \citep{ED6} calculates the Risk of Exposure-Induced Death (REID) as: 

\begin{equation}
    \mbox{REID}_c(e,D) = \int_e^\infty h_c(a|e,D) S^*(a|e,D) da,
\end{equation}
where $h_c(a|e, D)$ is the excess mortality rate due to cause $c$ and $S^*(a|e,D)$ is the survival function at age $a$ attributable to radiation exposure with dose $D$ at age $e$. However, in 2021, following the recommendations of the US National Academy of Sciences (NAS), NASA adopted a sex- and age-independent limit obtained by generalising the most susceptible case (30-year-old female) from the NSCR 2012 model and establishing a global limit of 600 mSv \citep{ED7}. 

Other space agencies have developed other models to define career limits, such as the Radiation-Attributed Decrease of Survival (RADS) from the European Space Agency (ESA), the Disability Adjusted Life Year (DALY) from the Canadian Space Agency (CSA), the Lifetime Cancer Mortality (LCM) from the Japan Aerospace Exploration Agency (JAXA) or Russian space agency (RSA) model. For stochastic effects, JAXA is the only space agency that still uses age- and sex-dependent limits, as shown in Table \ref{table:LimitSA}. 

\begin{table}[h!]
\begin{center}
\begin{tabular}{ c c c }
  \hline
 \multicolumn{3}{c}{\small \textbf{NASA}} \\
 \hline
  \multicolumn{3}{c}{\footnotesize 0.60 Sv} \\
  \hline
   \hline
  \multicolumn{3}{c}{\small \textbf{ESA, RSA, CSA}} \\
  \hline
  \multicolumn{3}{c}{\footnotesize 1 Sv} \\
  \hline
   \hline
  \multicolumn{3}{c}{\small \textbf{JAXA}} \\
  \hline
  \footnotesize Age at First Space Flight (years) &  \footnotesize Males (Sv) & \footnotesize Females (Sv) \\
    \hline
  \footnotesize 27 - 30 & \footnotesize 0.60 & \footnotesize 0.50 \\
  \footnotesize 31-35   & \footnotesize 0.70 & \footnotesize 0.60 \\
  \footnotesize 36-40   & \footnotesize 0.80 & \footnotesize 0.65 \\
  \footnotesize 41-45   & \footnotesize 0.95 & \footnotesize 0.75 \\
  \footnotesize $>$45   & \footnotesize 1.00 & \footnotesize 0.80 \\
  \hline
\end{tabular}
\end{center}
\caption{Effective-dose career limits established by different space agencies \citep{ED4,ED7,ED9} .}
\label{table:LimitSA}
\end{table}

In this work, a total effective dose equivalent of $343.6 \pm 2.4$ $ \mu$Sv/day was estimated to be received by a human body present on the surface of Mars. If no substantial changes occur in the primary GCR spectra, this quantity is equivalent to an effective dose equivalent of $125.4 \pm 0.9$ mSv received in a full year. 

Several types of Mars mission plans have been proposed according to missions' duration \citep{ED10}. In this work, a short-stay Mars mission is considered to consist of 620 days in free space and 30 days on Mars’ surface, while a long stay consists of an 18-month travel time from Earth to Mars and about 500 days at Mars. Considering the astronaut dose limits imposed by space agencies presented in Table \ref{table:LimitSA}, the 1 Sv career limit adopted by most space agencies, e.g., ESA, RSA, and CSA, would not be reached according to these calculations. However, the global NASA limit of 600 mSv and the JAXA limits of 600 mSv and 500 mSv for younger than 30 male and female astronauts, respectively, raise some concerns due to the effective dose equivalent proximity to these limits.

Furthermore, it is significant to note that when data from atomic bomb survivors are used to assess mortality for radiation danger to astronauts in space, risk estimations have relatively high uncertainties. Qualitative distinctions in the biological effects of $\gamma$-rays and heavy ions raise several doubts about the accuracy of any scaling approach. Like so, Martian missions are expected to easily exceed the current dose limits set for Low Earth Orbit (LEO) missions. Dose limits have not yet been defined for Mars missions, once new knowledge is needed for better risk assessment as well as further consideration on acceptable risk levels for such missions.

Table \ref{table:Missions} displays the most relevant dose estimates by records on passive dosimeters, tissue absorption and average quality factors estimates from flight spectrometers and radiation transport codes, registered on NASA crews for missions until 2004 \citep{ED3}. The $343.6 \pm 2.4$ $\mu$Sv/day calculated in this work for the Martian surface are similar to International Space Station (ISS) missions and some U.S. Space Shuttles flights — officially referred to as the Space Transportation System (STS). Thus, a mission on the surface of the Red Planet would not be much different than other already performed NASA missions, if performed under solar modulation parameters similar to the ones occuring from November 15, 2015, to January 15, 2016, time frame.

\begin{table}[t!]
\begin{center}
\begin{tabular}{ c c  c  c  c  c } 
  \hline
\multirow{2}{*}{ \small NASA Program}&  \small No.   &  \small $D$                     &  \small $E$     &      \small $D_T$    &        \small $E_T$\\
                                   &  \small Crews & \footnotesize (mGy/day) &  \footnotesize (mSv/day) &  \footnotesize (mGy)  &  \footnotesize (mSv) \\       
 \hline
  \footnotesize Mercury                       & \footnotesize 6     & \footnotesize 0.3   &  \footnotesize 0.55  & \footnotesize 0.1  & \footnotesize 0.15     \\
  \footnotesize Gemini                        & \footnotesize 20    & \footnotesize 0.49  &  \footnotesize 0.87  & \footnotesize 1.3  & \footnotesize 2.2   \\
  \footnotesize Apollo                        & \footnotesize 33    & \footnotesize 0.43  &  \footnotesize 1.2   & \footnotesize 4.1  & \footnotesize 12   \\
  \multirow{2}{*} \footnotesize Skylab                      & \multirow{2}{*} {\footnotesize 9 } & \multirow{2}{*} {\footnotesize \footnotesize 0.71 } & \multirow{2}{*} {\footnotesize \footnotesize 1.4 } & \multirow{2}{*} {\footnotesize \footnotesize 40.3 } & \multirow{2}{*} {\footnotesize \footnotesize 95 } \\
  \footnotesize (50 deg, 430 km)\\
  \multirow{2}{*} \footnotesize ASTP                      & \multirow{2}{*} {\footnotesize 3 } & \multirow{2}{*} {\footnotesize \footnotesize 0.12 } & \multirow{2}{*} {\footnotesize \footnotesize 0.26 } & \multirow{2}{*} {\footnotesize \footnotesize 1.1 } & \multirow{2}{*} {\footnotesize \footnotesize 2.3 } \\
  \footnotesize (50 deg, 220 km)\\
  
  \footnotesize STS    \\
  
  \footnotesize (28.5 deg, $>400$ km) & {\footnotesize 85 } &  {\footnotesize \footnotesize 1.2 } &  { \footnotesize 2.1 } &  {\footnotesize  9.5 } & { \footnotesize 17 } \\
  \footnotesize (28.5 deg, $<400$ km) & {\footnotesize 207 } & {\footnotesize \footnotesize 0.1 } &  {\footnotesize 0.18 } &  {\footnotesize 0.9 } &  { \footnotesize 1.6 } \\
  \footnotesize (39 - 40 deg) &  {\footnotesize 57 } &{\footnotesize \footnotesize 0.1 } & {\footnotesize \footnotesize 0.21 } & {\footnotesize \footnotesize 1.1 } & {\footnotesize \footnotesize 2.4 } \\
  \footnotesize ($>50$ deg, $>400$ km) &  {\footnotesize 10 } &  {\footnotesize \footnotesize 0.44 } &  {\footnotesize \footnotesize 1.1 } & {\footnotesize \footnotesize 2.2 } & {\footnotesize \footnotesize 5.2 } \\
  \footnotesize ($>50$ deg, $<400$ km) &  {\footnotesize 190 } &  {\footnotesize \footnotesize 0.2 } &  {\footnotesize \footnotesize 0.45 } &  {\footnotesize \footnotesize 1.7 } &  {\footnotesize \footnotesize 3.8 } \\ 
  \multirow{2}{*} \footnotesize NASA-Mir                      & \multirow{2}{*} {\footnotesize 6 } & \multirow{2}{*} {\footnotesize \footnotesize 0.37 } & \multirow{2}{*} {\footnotesize \footnotesize 0.84 } & \multirow{2}{*} {\footnotesize \footnotesize 50.3 } & \multirow{2}{*} {\footnotesize \footnotesize 115} \\
  \footnotesize (51.6 deg, 360 km)\\ 
  \multirow{2}{*} \footnotesize ISS                      & \multirow{2}{*} {\footnotesize 13 } & \multirow{2}{*} {\footnotesize \footnotesize 0.16 } & \multirow{2}{*} {\footnotesize \footnotesize 0.4 } & \multirow{2}{*} {\footnotesize \footnotesize 26 } & \multirow{2}{*} {\footnotesize \footnotesize 68} \\
  \footnotesize (51.6 deg, 380 km)\\  
  \hline
\end{tabular}
\end{center}
\caption{Average dose rate and total dose recorded by dosimetry badges and effective dose estimates received by  NASA program crews until 2004 \citep{ED3}.}
\label{table:Missions}
\end{table}

Although not considered in this work, the occurrence of Solar Particle Events (SPE) and exposure to ionising radiation in transit to Mars are scena\-rios not to be neglected due to the high dose rates they can cause, easily exceeding the career limits and causing deterministic effects in a human body. On November 26, 2011, the Curiosity rover started its 253-day journey to Mars. During the cruise, the MSL-RAD instrument performed measurements of the radiation environment inside the spacecraft \citep{ED11}. The average GCR absorbed dose rate measured in RAD’s plastic scintillator detectors (composition similar to that of human tissue) during quiet periods in solar activity, was about $461 \pm 92$ $\mu$Gy/day, a twofold amount from the one measured at the Martian surface ($233 \pm 12$ $\mu$Gy/day). During the few SPE detected, measured dose rates were as high as $>$10 000 $\mu$Gy/day.

\section{Conclusion}
\label{sec:Conclusion}

The research here presented performed simulations using the state-of-art three-dimensional MCNP6 Monte Carlo particle transport code to investigate the interactions of space radiation with the Martian environment. The viability of astronauts staying on the Martian surface for extended periods of time was assessed, by determining dose rate magnitudes from GCR exposure and comparing them with effective-dose career limits. 

MCNP6 revealed to be a good code for high-energy GCR transport in the Martian environment as evidenced by the consistent secondary spectra agreement with other transport codes and RAD measurements. Dose rates are in line with those reported by RAD, with the exception of protons, to which an overestimation of the deposited energy occurs in the MCNP6.2 version calculations. Corrections using the ICRP’s particle fluence to organ absorbed dose conversion coefficients from Publication 123 were employed to proton’s values. 

During the November 15, 2015, to January 15, 2016, period, GCR were at an intermediate intensity after a minimum had been reached during a relatively weak solar activity maximum at the end of 2014. From GCR exposure, an effective dose equivalent of $343.6 \pm 2.4$ $\mu$Sv/day would be received by an astronaut, which is equivalent to $125.4 \pm 0.9$ mSv received during a full year, considering constant primary GCR spectra for that period. Protons, neutrons, and heavier ions (2 $\leq$ Z $\leq$ 28) would be the main contributors to the total effective dose equivalent. 

Missions’ duration will play an important and limiting factor for mission viability. For a short Mars mission, the REID would be lower than the required 3\%, but for a long-stay mission, effective dose equivalent values could reach career limits established by space agencies. Important to note is that current limits are defined for LEO missions, and that for deep space missions, space agencies are currently working to define appropriate new limits. 


Radiation risk is immeasurable by definition. Therefore, it must be understood using a combination of basic biological damage information and human radio-epidemiology data. Further research and new knowledge are needed before confidence in risk projections for Mars missions is improved to a satisfactory level.

\section*{Acknowledgements}
This work was supported by FCT - Fundação para a Ciência e a Tecnologia through national funds and by FEDER through COMPETE2020 - Programa Operacional Competitividade e Internacionalização by these grants: UIDB/04434/2020; UIDP/04434/2020.

\bibliographystyle{elsarticle-harv} 
\bibliography{cas-refs}





\end{document}